\documentstyle[multicol,aps,prb,epsf]{revtex}
\begin{document}
\pagestyle{empty}
\newcommand{\bc}{\begin{center}}
\newcommand{\ec}{\end{center}}
\newcommand{\be}{\begin{equation}}
\newcommand{\ee}{\end{equation}}
\newcommand{\beqn}{\begin{eqnarray}}
\newcommand{\eeqn}{\end{eqnarray}}

\begin{multicols}{2}
\narrowtext
\parskip=0cm

\noindent
{\large\bf Comment on "The dead zone for string players"}  
\smallskip

In a recent Letter Broomfield and Leask \cite{BL} experimentally found that
"It is not possible to produce sound from a string when it is
bowed at its midpoint."
This finding is trivial if one knows the following mechanism of bowing:

It is well known\cite{MORE} that if one puts a  horse-tail bow hair
under a microscope, there are many 'small nails' on its surface. 
It is these 'nails' which continuously
pluck the string as the bow moves that produce sound.
This is different from what happens when one plucks
a string with one's finger. In the latter case the string can vibrate
freely once the finger has left the string,
while in the former case the fingers constantly intervenes the vibration. 
The nail spacing is so close,
perhaps it is better to describe these 'nails' as 'brushes' or 'combs'.
This kind of vibration is therefore 
better described as a 'forced' or 'kicked' vibration.
Once one nail has passed over the string, 
the string can only vibrate freely for a short period before the next
nail arrives. 
It is therefore better for the amplitude of vibration at the bowing point to
be about the same as the spacing of the nails so that the intervention of the
nails is minimized. If the amplitude is larger than the spacing of the nails,
the string does not manage to vibrate much before the next nail arrives.

There are already mass-and-spring models for bowing of string
instruments in the literature\cite{NHF}.
Generally, they consider bowing as a {\it macroscopic} frictional effect.
These friction-like models will have difficulty to explain the silence
of bowing in the middle of the string.
However, with the {\it microscopic} mechanism proposed  here the phenomenon
is clear.

This mechanism applies to every part of the string, but the influence
of the next nail becomes most significant as one bowes at the
midpoint. 
Furthermore, the mechanism has nothing to do with the resin used.
Resin increases the friction, or the length and the number of the nails.
A clean bow without any resin can still produce sound.

More refined models can be developed
based on the reasoning given above.
It will be interesting to find out experimentally what kind of modes are 
excited at different bowing conditions. 
Remember that the string is not a rigid body and the nails do not always pluck
the string in a direction normal to it.          
Therefore, the generation of the vibration and the
interference of the next nail is always imperfect.
In the field of nonlinear dynamics, phenomena 
of 'kicked oscillators', which
display interesting chaotic behaviour, have been studied.
It will also be interesting to see how they apply to the bowing of
violins.

\bigskip
\noindent
{Julian Juhi-Lian Ting}

{\small \noindent
jlting@multimania.com
}

\bigskip
\noindent
Date: \today

\begin{figure}
\epsfxsize=8cm\epsfbox{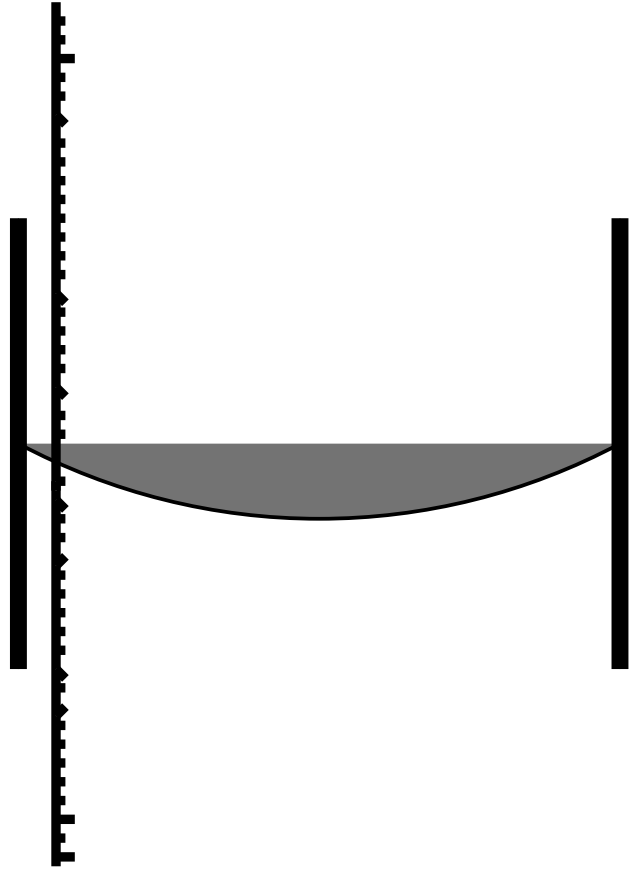}
\vspace{0.3cm}
\caption{Schematic plot of bow-hair string interaction.
Bowing a string is just like sawing it. 
The sound generating mechanism is something in-between 
plucking a string with one's finger and 
microscopical frictional effects.
One should consider the process as a kicked vibration.
The presence of the next nail will stop the 
vibration if one bows at the middle of the string.
}
\end{figure}

\end{multicols}

\end{document}